# Single-Mode Emission in InP Microdisks on Si using Au Antenna

*Preksha Tiwari, Anna Fischer, Markus Scherrer, Daniele Caimi, Heinz Schmid, and Kirsten E. Moselund\**

IBM Research Europe - Zurich, Säumerstr. 4, 8803 Rüschlikon, Switzerland

**\*Correspondence to: kmo@zurich.ibm.com**

**Abstract**

**An important building block for on-chip photonic applications is a scaled emitter. Whispering gallery mode cavities based on III-Vs on Si allow for small device footprints and lasing with low thresholds. However, multimodal emission and wavelength stability over a wider range of temperature can be challenging. Here, we explore the use of Au nanorods on InP whispering gallery mode lasers on Si for single mode emission. We show that by proper choice of the antenna size and positioning, we can suppress the side-modes of a cavity and achieve single mode emission over a wide excitation range. We establish emission trends by varying the size of the antenna and show that the far-field radiation pattern differs significantly for devices with and without antenna. Furthermore, the antenna-induced single mode emission is dominant from room temperature (300 K) down to 200 K, whereas the cavity without an antenna is multimodal and its dominant emission wavelength is highly temperature dependent.**

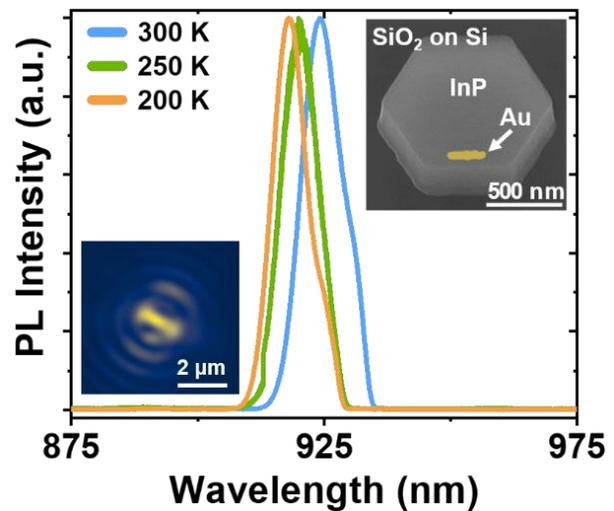

## Introduction

Integrated light sources show potential for a wide range of applications, from optical communication to quantum information processing to sensing. Using Si as a material platform allows to leverage established fabrication processes for passive structures. Due to its indirect bandgap, however, an alternative material is needed for active devices. Group III-V semiconductors pose a viable choice for emitters [1–5] and detectors [6–11] due to their direct and tuneable bandgap, high mobilities, and high absorption coefficients covering the entire telecommunication band.

Small mode volumes and low thresholds can be achieved by various cavity types, like photonic crystal cavities [12–17], metal-clad cavities [18–23], semiconductor-on-metal cavities [24,25], or whispering gallery mode microdisk cavities [26–29] based on total internal reflection. The latter have the advantage of possessing a simple fabrication scheme. However, mode selectivity and multimodal emission can be a challenge, which will be addressed in the present work. Furthermore, a general challenge in III-V semiconductors photonics is their strong temperature sensitivity: The bandgap follows the Varshni shift [30], leading to a change in spectral overlap between the material gain and resonant mode wavelengths at different temperatures.

Single mode emission can be achieved by supressing side-modes and breaking the symmetry of the devices: One way to achieve single mode lasing in microdisk cavities was demonstrated using suspended cavities with a proper choice of bridges manipulating the spatial symmetry [31]. Another strategy involves using grooves [32] or nanoantennae [33,34]. Recently, nanoantennae have also been combined with high-Q cavities containing quantum dots, leading to hybrid systems where emission enhancements exceed those of a bare cavity and allow for tuning the bandwidth [35,36]. These demonstrations show the strong potential of nanoantennae for single-photon devices [37] and towards strongly coupled systems [38]. For plasmonic nanoantennae it has been shown that, following the Mie-Gans scattering, the scattering cross sections in the visible and near infrared as well as resonances are tunable and dependent on the aspect ratio of the antennae [39–41]. This enables optimization of antenna geometry to enhance or suppress emission of a specific wavelength range.

While side-mode suppression and enhanced directivity were successfully demonstrated for microdisk cavities coupled with a Pt antenna without degradation of the dominant mode in terms of threshold [33], the impact of the metallic antenna and its geometry on wavelength stability, also considering different temperatures, remains to be studied. Here we explore the effect of Au nanorod antennae on top of InP microdisk lasers fabricated on Si. Using a relatively simple process based on direct wafer bonding, etching, and lift-off, we are able to fabricate a large number of devices. This allows us to get insight into general trends of the antenna size and position on the resonant emission of the WGM cavities. We observe significantly improved device performance in terms of side-mode suppression and wavelength stability for different temperatures. We believe that these findings are of general interest for the optimization of the emission characteristics of micro and nanolasers.

**Device Fabrication**

A 225 nm thick InP layer is grown on a lattice-matched sacrificial InP wafer with an InGaAs etch stop layer in between using metal-organic chemical vapor deposition (MOCVD). Then, the material is bonded onto a Si wafer with a 2 µm thick $SiO_2$ layer in between, serving as an optical insulator layer, and the donor wafer material is removed. More information on direct wafer bonding techniques can be found in [42]. For the microdisk cavities, the antennae are first defined by a lift-off process using a PMMA bilayer as resist and 40 nm electron-beam evaporated Au and a 2 nm Ti adhesion layer. Hexagonal microdisks are then patterned using HSQ as a resist. InP microdisks with a width of 1100 nm and a thickness of 225 nm are etched by inductively coupled plasma (ICP) dry etching using $CH_4$, $Cl_2$ and $H_2$ . After the etch, the sample is cleaned with a 1:10 diluted phosphoric acid solution and capped with 3 nm of $Al_2O_3$ using atomic layer deposition. The antennae are between 40 nm to 70 nm wide and 150 nm to 300 nm long and are either placed along (parallel) the side facet of the InP cavity or rotated in-plane by 90 degrees (orthogonal) with respect to it. The distance between the antenna and the cavity edge is designed to be 50 nm but varies due to drift and alignment accuracy during the patterning process or non-optimal adhesion. Fig. 1 illustrates the fabrication steps and shows a SEM image of the final device with a parallelly placed antenna.

All measurements are performed with a micro-photoluminescence (micro-PL) setup where a ps-pulsed excitation source with emission at 750 nm and a repetition rate of 78 MHz is focused onto the device with a 100 x objective (NA = 0.6) and a spot-size of approximately 1 µm. The emitted photoluminescence is collected in reflection mode from the top of the device and the spectrum is detected by a linear array InGaAs detector. In the following, we will discuss the impact on the emission spectrum of the microdisk through the use of Au antennae with varying cross-sectional area. In some cases, the different antenna widths will be additionally color coded in the figures, in order to map them to respective cross-section areas.

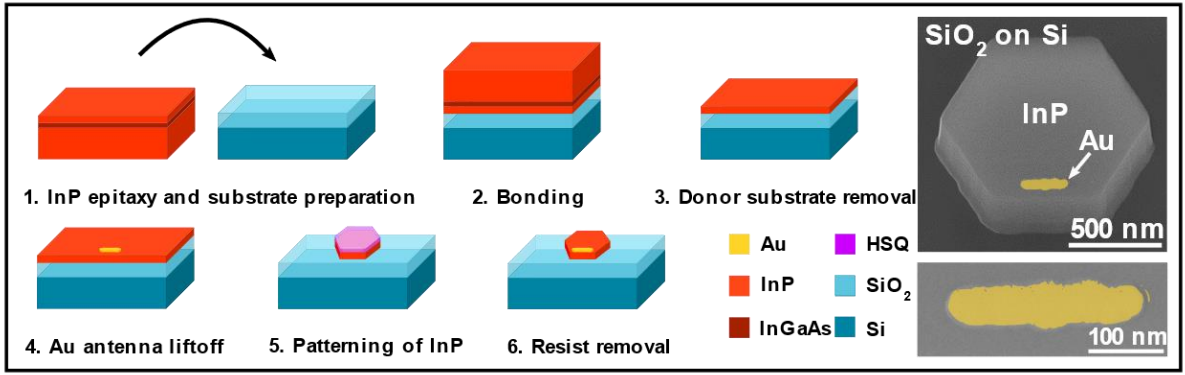

*Figure 1.* Illustration of the fabrication steps. After wafer bonding, the Au antennae are defined via lift-off. The InP microdisk cavities are etched afterwards and the top HSQ resist is removed. The SEM image shows a device with a parallelly placed antenna, and a zoom-in of the Au nanorod (false-colored).

**Comparison of Different Orientations**

Fig. 2(a) shows the PL emission spectra of a 1.1 µm wide device without antenna upon increasing input power with the bulk emission of InP in the inset. Two resonant emission peaks at around 925 nm (peak 1) and 960 nm (peak 2) emerge at higher excitation energies. Fig. 2(b) shows the emission spectra upon increasing input power for a cavity with a parallelly oriented Au antenna on top. Compared to the bare cavity case, peak 2 is supressed.

The light-in-light-out (LL) curve in Fig. 2(c) shows a multimode behavior with similar thresholds for peak 1 and 2 in the bare cavity case, which are 0.8 pJ/pulse and 1.2 pJ/pulse respectively. For the antenna-coupled cavity, the threshold of peak 1 is comparable to those of the bare cavity (1.1 pJ/pulse) and its intensity is slightly higher, but the latter is most likely a result of the stronger emission of InP at a wavelength of 925 nm compared to 960 nm. Peak 2 of this device, however, is significantly suppressed and only appears at higher powers, whereas at pump powers below ~6 pJ/pulse the antenna-coupled device is single mode, i.e., no second resonant wavelength peak is visible.

From LL curves like the one in Fig. 2(c), peak ratios between peak 1 and peak 2 are determined and illustrated in Fig. 2(d) for the bare cavity and four different antennae of varying dimension and position: For the bare cavity, peak 1 is the dominant peak at first, then the ratio between the amplitude of the two peaks rapidly decreases and the longer-wavelength peak 2 dominates after around 2 pJ/pulse, indicated by a peak ratio value < 1. On the other hand, for the parallelly oriented antennae, peak 1 stays dominant for the larger antennae over the entire excitation range. Only the small antenna with a cross section of 0.007 µm$^2$ has a crossover at around 10 pJ/pulse. Also, it is notable that the antenna-coupled devices are single-mode for lower excitation powers, and the slight multimode behavior only appears for increasingly higher powers as the size of the antenna is decreased.

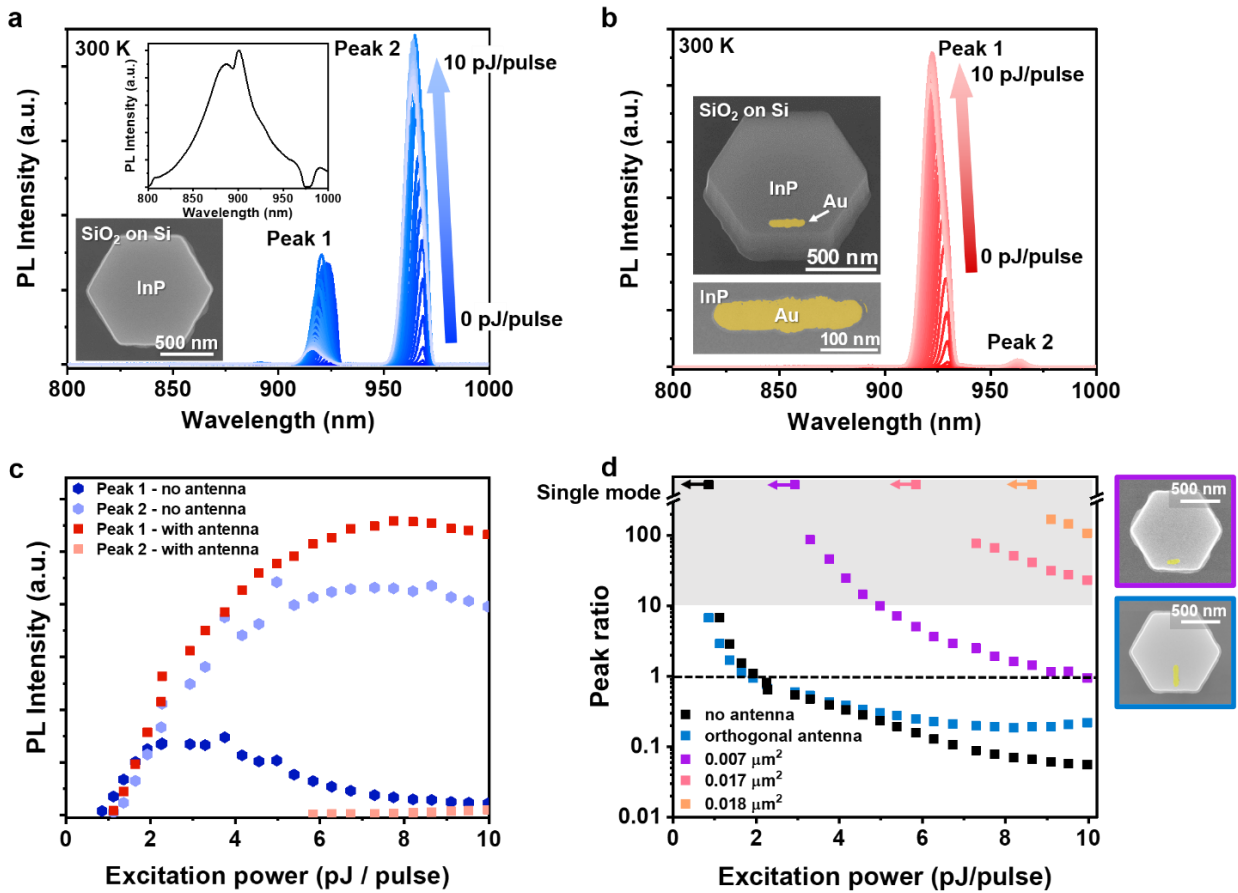

*Figure 2. (a) Bulk emission (inset) and emission spectrum upon increasing excitation energy for a 1100 nm wide and 225 nm thick InP cavity bonded on Si. (b) Emission spectrum for a 1100 nm wide InP cavity with a parallelly oriented Au antenna on top. (c) Light-in-light-out curve for the spectrum shown in (a) and (b) from which the threshold is extracted. (d) Peak ratio of peak 1 at 925 nm and peak 2 at 960 nm for a bare cavity and antenna with varying sizes and orientations. The shaded part highlights the region where the peak ratio is greater than 10. The points with arrows pointing towards the "single mode" label, mark the excitation powers, up to which a certain device is single mode, i.e., where only one resonant emission peak is visible. Corresponding SEM image for the orthogonal antenna and the parallel antenna with an area of 0.007 μm² are shown on the right.*

For the orthogonally placed antenna, the multi-mode behavior is similar to the bare cavity and there is no side-mode suppression effect. Only at higher excitation powers does the peak ratio deviate, potentially due to mode competition and other effects in the cavity. The low selectivity of the orthogonally placed antenna can potentially be attributed to the lower overlap of the resonant mode, polarization of the mode, and the scattering cross section of the antenna: The whispering gallery mode is expected to be at the periphery of the cavity; hence, an orthogonal placed antenna may overlap with a node of the mode or only partially if not placed accurately. For the parallely placed nanoantenna, exact positioning may be less crucial because a larger fraction of the antenna is expected to be at the position of the electric field of the resonant mode.

Far-field radiation images showing first-order interference patterns were captured with an infrared camera and are shown in Figure 3(a) in the case of a bare cavity and in Figure 3(b)-(e) of representative antenna-coupled structures. While the far-field has a circular shape for the bare cavity case, it is dumbbell-shaped for the devices with an antenna. This suggests that the dominant resonant mode does interact with the antenna which acts as a near-to-far-field coupler.

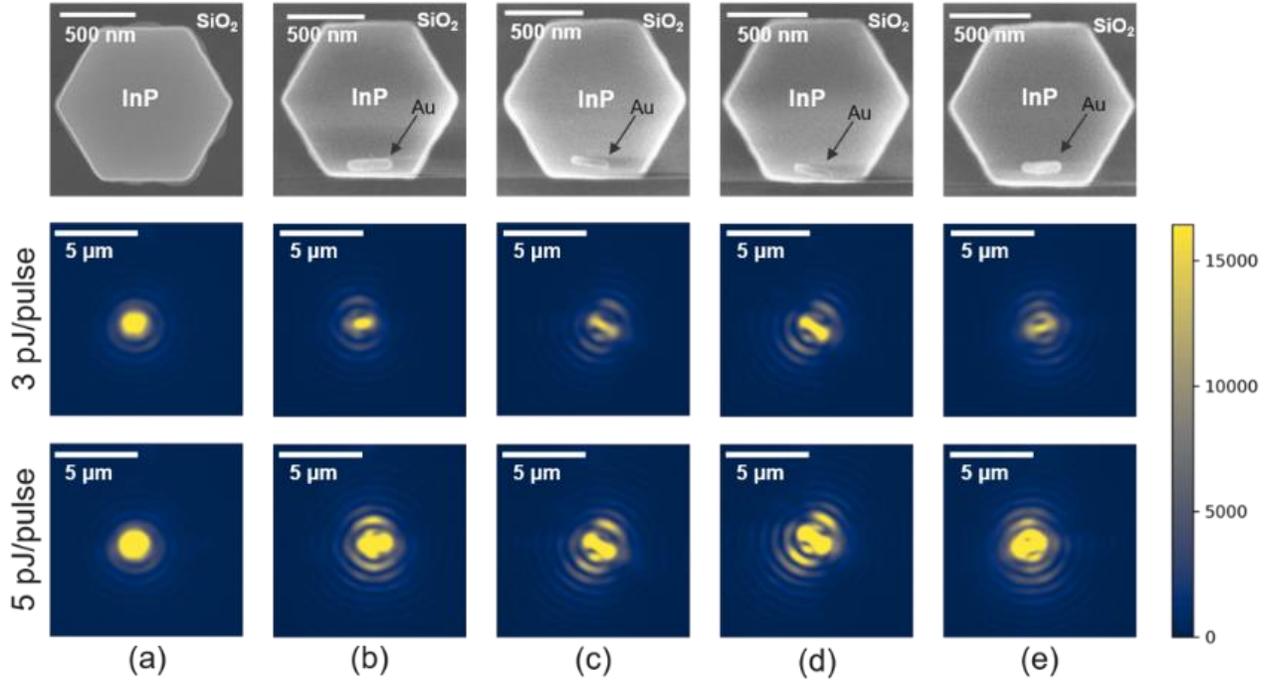

*Figure 3. Far-field radiation images with first-order interference patterns for devices (a) without and (b) - (e) with antenna at 3 pJ/pulse and 5 pJ/pulse.*

**Selectivity for Different Antenna Areas**

In total, 121 devices with parallelly placed antennae covering areas from 0.00605 µm$^2$ to 0.0194 µm$^2$ were measured and the following results shall be representative for general trends. Fig. 4 shows the peak ratio versus the antenna cross-section area at 4 pJ/pulse and 10 pJ/pulse for all the measured devices. The peak ratios are additionally color coded, corresponding to the different antenna widths for clearer visualization. If only one emission peak is visible, we refer to the devices as single mode. We define peak 1 to be dominating if the peak ratio rises over 10. When the peak ratio is below 1 this means that peak 2 is the dominant peak. This is the case for cavities without antenna (see solid line in Fig. 4 which corresponds to peak ratio values for the bare cavity), or for some of the smallest antennae sizes, especially at higher excitation energies. In general, antenna-coupled devices show a higher peak ratio value with a more dominant emission of peak 1 for the entire excitation rage. Some cavities even exhibit single mode emission up to 10 pJ/pulse. Furthermore, the larger antennae tend to be more selective than the smaller ones with higher peak ratios and more single mode devices. The overall spread in the peak ratio for a certain antenna area can be explained by fabrication-related deviations of the antenna shape (varying width across the structure) and positioning (slight tilt and different edge-to-antenna spacing due to adhesion and drift during processing, see Fig. 3).

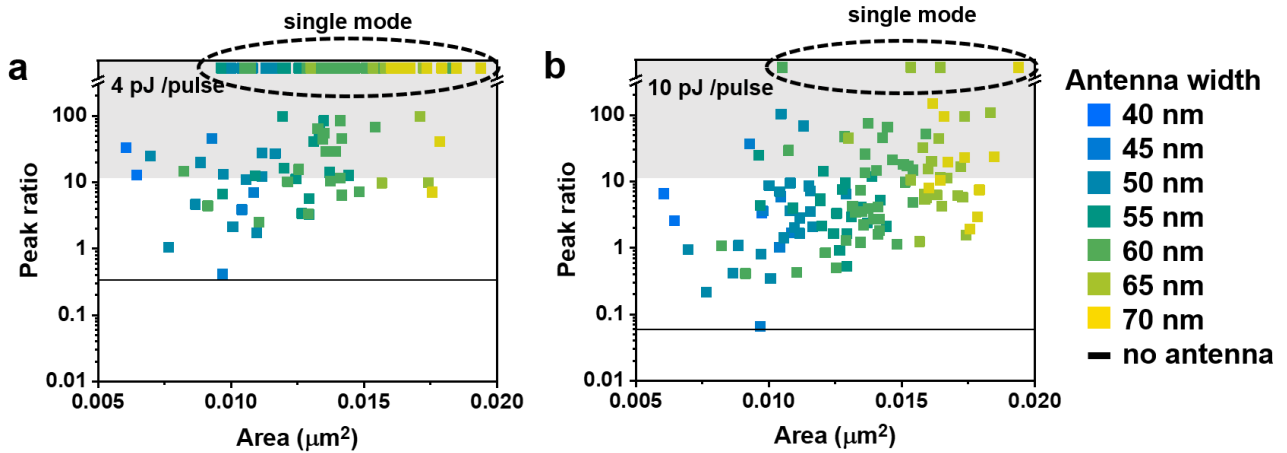

*Figure 4. Peak ratio (counts peak 1/counts peak 2) for all the measured devices versus antenna area at an excitation power of (a) 4 pJ/pulse, and (b) 10 pJ/pulse. The circled devices were single mode (only one resonant emission peak visible). The shaded area corresponds to peak ratios which are higher than 10.*

To quantify this trend further, the devices were binned into quintiles ranging from the smallest (Q1) to the largest (Q5) device areas and the number of single mode devices at different powers were determined. Figure 5(a) shows the percentage of measured devices which were single mode at 4 pJ/pulse, 6 pJ/pulse, 8 pJ/pulse and 10 pJ/pulse. Figure 5(b) shows the absolute number of measured and single mode devices corresponding to the percentage shown in Figure 5(a). As in Figure 4, a trend is visible towards larger antenna areas: While in the smallest quintile no device was single mode, in the largest two there are 80% and almost 70% respectively. At 10 pJ/pulse only 4 of the measured devices are single mode, and three of those are in the largest two quintiles. It seems that the most selective antennae were in the bin Q4, corresponding to the second largest quintile with areas ranging from 0.0141 µm$^2$ – 0.0167 µm$^2$. It should be noted, however, that for Q5 the total number of measured devices is lower than for Q4 (12 versus 35 devices), so a less selective antenna will weigh heavier in the percentage.

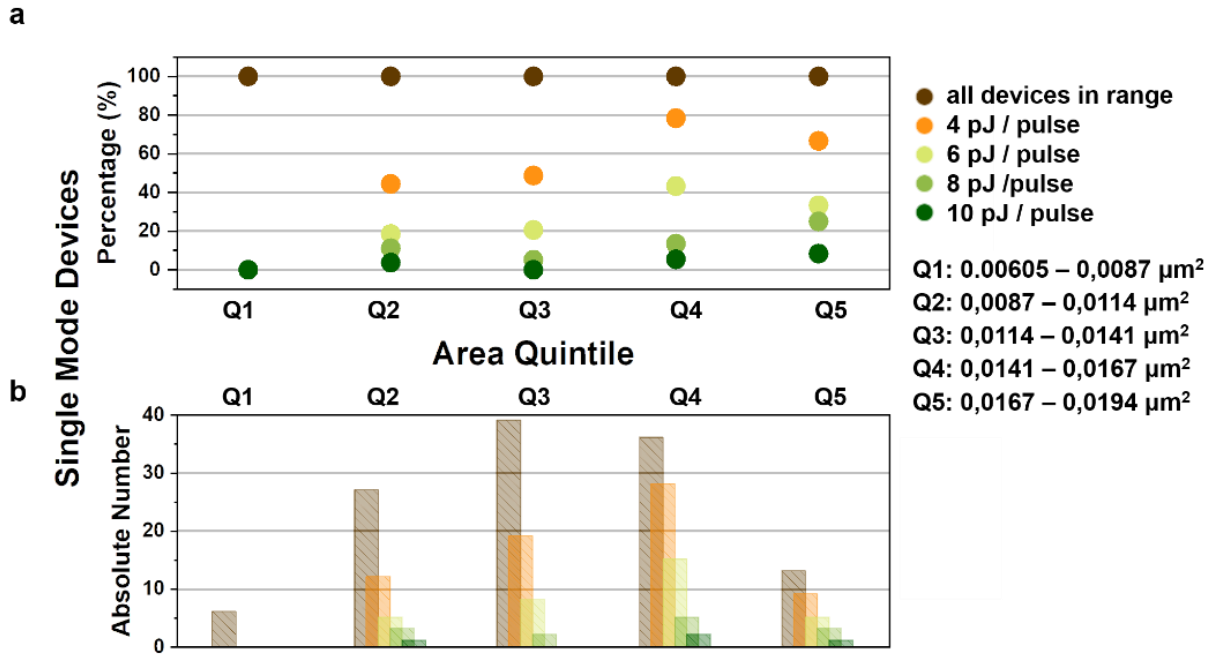

*Figure 5. Single mode devices binned into areas of smallest to largest quintile, (a) in percentage and (b) in absolute numbers.*

These results may be taken as an indication that there is a certain optimum in terms on antenna size: Initially, an increase in antenna size will provide improved side-mode suppression, whereas eventually this effect saturates. It is expected that several effects might impact the mode selectivity and device performance: The scattering and absorption cross sections at a certain wavelength depend on the antenna area and substrate [39,40,43,44]. So, depending on the antenna size and position, one resonant mode may be subject to stronger absorption than the other, or both modes might be affected similarly.

**Performance at Different Temperatures**

To investigate the extent to which the antenna not only allows for side-mode suppression but also wavelength stability, we performed micro-PL measurements at various temperatures. Since the bandgap of InP is temperature-dependent, the gain emission peak shifts to lower wavelengths upon temperature decrease, leading to a change of the dominant resonant mode to the one which now has a stronger overlap with the bulk PL. In the antenna-coupled case, however, the resonant wavelength stays the same down to 200 K, as it is shown in Fig. 6(a). This indicates that the antenna is not merely preferentially scattering a particular wavelength, but that it enhances either the dominant mode or supresses the others and thereby counterbalances the temperature-dependent shift of the gain. This is a significant result as temperature stabilization in nanophotonic components is a great challenge.

When placing a metal in close proximity to a resonant cavity, the question naturally arises whether this will lead to an increase in absorption losses and thereby an increase of the threshold for resonant emission. Fig. 6(b) shows the threshold of the different devices versus antenna area for peak 1 at 300 K. The square, colored points correspond to devices, which are single mode at least up to 6 pJ/pulse and the grey points correspond to all the antenna-devices which were measured. The threshold of peak 1 is higher for the devices which were single mode up to 6 pJ/pulse and comparable to the bare cavity case for the other devices. This indicates that the most effective antenna in terms of side-mode suppression leads to a higher threshold, likely due to increased absorption losses associated with the antenna and the optical mode which is directly disturbed by it.

Fig. 6(c) shows the relative blue shift of the different devices which were single mode up to 6 pJ/pulse and for the bare cavity case. The blue shift was measured at power levels which corresponded to 2x (filled symbols for antenna-coupled devices) and 4x (empty symbols for antenna-coupled devices) the threshold. The solid lines correspond to the average blue shift of 10 bare cavities at 2x (black) and 4x (grey) the threshold. They are around 1.5 nm at 2x threshold and almost 6 nm at 4x threshold power. The dashed lines show the standard deviation of the blue shift for the bare cavities. The blue shift of the resonant mode upon increasing input power is related to the plasma dispersion effect [45], a change in refractive index caused by the presence of free carriers, and is commonly observed in III-V semiconductor lasers [18,26,46]. Interestingly, the blue shift at 2x the threshold is in a comparable range for the different kind of devices, but it is larger for the bare cavity case than for the antenna-coupled devices at 4x the threshold. This indicates that the antenna effectively clamps the emission wavelength of the resonant mode. This would support the assumption that the mode selectivity may result from a plasmonic effect.

An assessment of the carrier dynamics would be interesting, since plasmonics can for example be used for high-speed photonics components, such as detectors and modulators [47–49]. This was unfortunately not possible in the given setup, due to the resolution limit of the lifetime measurement set-up which is approximately 50 ps. Therefore, we cannot resolve the fast dynamics of this system.

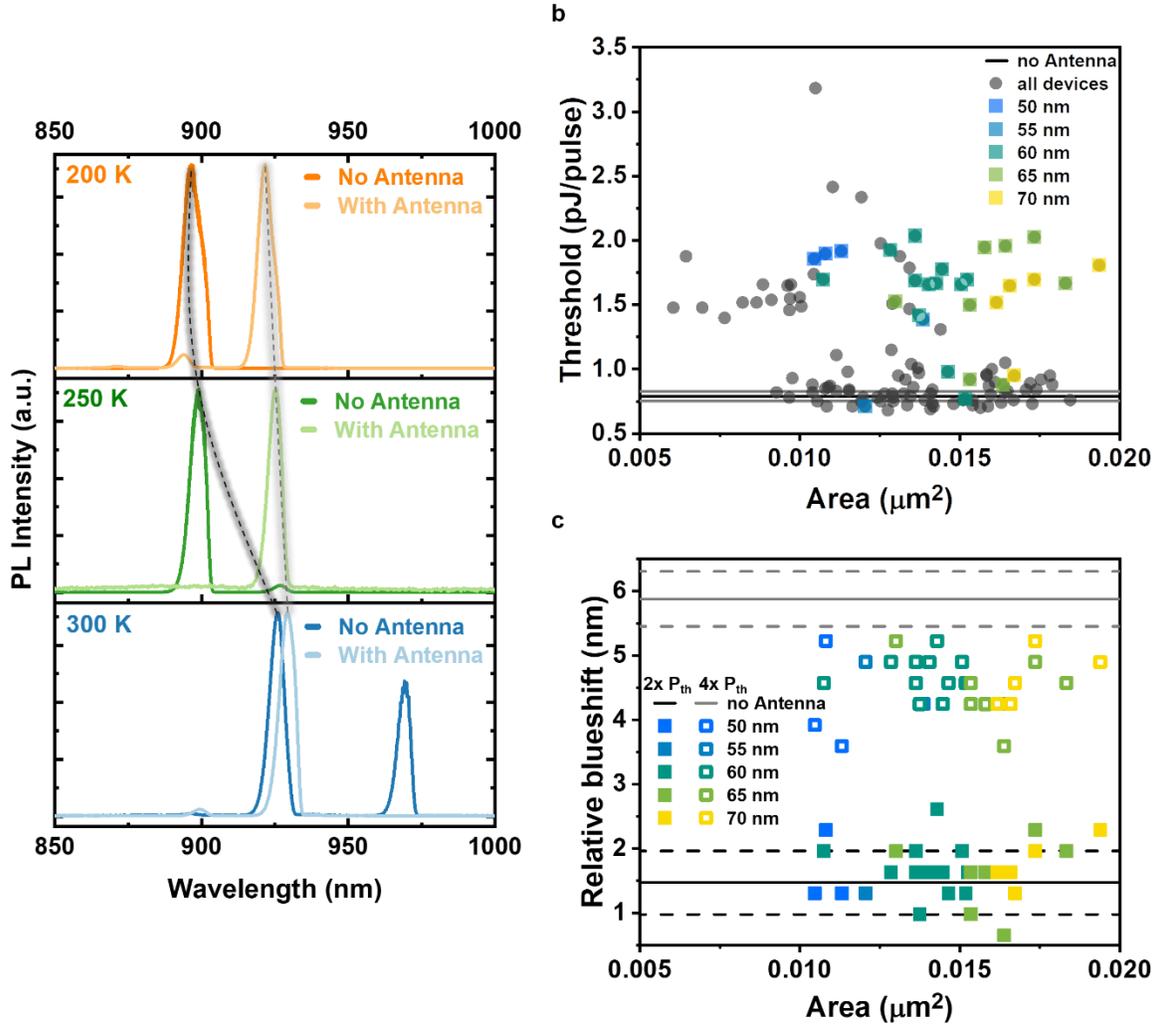

*Figure 6. (a) Spectrum for devices with (light colors) and without antenna (dark colors) at different temperatures. The dark dashed line outlines the change of the dominant emission mode for the bare cavity, the light dashed line tracks the peak position for the antenna-coupled devices. (b) Threshold of the bare cavity (solid black line for average of ten devices and grey line indicating the standard deviation) and antenna-coupled devices versus antenna area.. The square, colored points correspond to devices, which are single mode up to 6 pJ/pulse or more, and the grey points correspond to all the antenna-devices which were measured. (c) Relative peak shift for devices which were single mode up to 6 pJ/pulse or more at 2x (solid points), and 4x (hollow points) the threshold. The solid line corresponds to the average blue shift of 10 bare cavities (around 1.5 nm at 2x threshold and almost 6 nm at 4x threshold power). The dashed line shows the standard deviation of the blue shift for the bare cavities.*

**Conclusion and Outlook**

In this work, we presented a systematic study on the effect of Au nanorod antennae on InP whispering gallery mode cavities. While the bare InP cavity is multimodal and the dominant resonance wavelength changes significantly with temperature, we show that in antenna-coupled devices, we can achieve single mode emission and wavelength stability over 100 K (from 200 K up to room temperature at 300 K). The antenna must be aligned properly to the optical mode, as we only observed this side-mode suppression for antennae aligned along the cavity periphery (parallel) and not angularly (orthogonal). The beneficial effect of the antennae initially increases with its relative size, until it saturates at dimensions around $0.014\ \mu m^2 - 0.017\ \mu m^2$. The antenna reduces the relative blue shift due to the plasma dispersion effect, thereby providing for more stable emission at higher excitation powers. From the combination of these results, we can conclude that the antenna does not just impact light collection via scattering, but that the presence of the metal stabilizes the emission mode, thereby improving side-mode suppression and wavelength stability over variations in temperature and pumping powers.

Furthermore, we found that the threshold of the dominant peak is increased in antenna-coupled devices which are single mode up to 6 pJ/pulse, whereas devices which have a lower side-mode suppression are more likely to have similar thresholds compared to the devices without antenna. We attribute absorption losses due to the antenna to be the cause of the higher threshold.

Table 1 summarizes some parameters found in the experiments above. Listed are the average values for the bare cavity and for selected antenna-coupled devices, which were single mode up to 6 pJ/pulse. For the antenna-coupled devices, values corresponding to the device with the overall highest side-mode suppression (single mode at > 10 pJ/pulse), the lowest relative blue shift, and the lowest threshold are listed.

We believe the findings to be relevant and portable to abroad range of nanophotonic architectures. A long-term goal would be the coupling to more complex laser architectures or exploration of antenna shapes specific to desired emission wavelengths and directions.

|  | Antenna area [$\mu m^2$] | Threshold $P_{th}$ [pJ/pulse] | Blue shift at 2x $P_{th}$ [nm] | Blue shift at 4x $P_{th}$ [nm] | Peak ratio at 6 pJ/pulse |
|---|---|---|---|---|---|
| Bare cavity | - | 0.79 ± 0.04 | 1.5 ± 0.5 | 5.9 ± 0.4 | 0.15 ± 0.05 |
| Antenna with highest side-mode suppression | 0.0153 | 1.49 | 1.6 | 4.25 | Single mode |
| Antenna with lowest blue shift | 0.0164 | 0.87 | 0.65 | 3.6 | Single mode |
| Antenna with lowest threshold | 0.0121 | 0.7 | 1.3 | 4.9 | Single mode |

*Table 1. Comparison of bare cavities and selected antenna-coupled devices which were single mode (only peak 1) up to 6 pJ/pulse. Listed are average threshold, relative blue shift at 2x and 4x the threshold, and peak ratios at 6 pJ/pulse. For the antenna-coupled devices, the values for the microdisk with the highest side-mode suppression (single mode > 10 pJ/pulse), lowest blue shift, and lowest threshold are given.*

**Acknowledgments**
The authors gratefully acknowledge Rachel Grange and Andreas Schenk for fruitful technical discussions. We thank the Cleanroom Operations Team of the Binnig and Rohrer Nanotechnology Center (BRNC) for their help and support. The work presented here has received funding from the European Union H2020 ERC Starting Grant project PLASMIC (Grant Agreement No. 678567) and the European Union H2020 program SEQUENCE (Grant No.871764).

**Author Contributions**
P.T., A.F, and K.M. created the device concept. P.T. fabricated the sample with support of M.S. and H.S. for material growth, D.C. for the wafer bonding, and A.F. for lift off. A.F. and P.T. performed the optical characterization on the devices. A.F. and P.T. analyzed the data. K.M. lead and managed the project. All authors discussed the results. The manuscript was written by P.T and K.M., with contributions of all authors, and all authors have given approval to the final version of the manuscript.